\documentclass[12pt, letterpaper, preprint, comicneue]{aastex63}
\usepackage[T1]{fontenc}
\usepackage{fontawesome}
\usepackage{color}
\usepackage{amsmath}
\usepackage{natbib}
\usepackage{ctable}
\usepackage{bm}
\usepackage[normalem]{ulem} 
\usepackage{xspace}
\usepackage{paralist}
\usepackage{fontawesome}
\usepackage{multirow}

\let\oldbibliography\thebibliography 
\renewcommand{\thebibliography}[1]{%
  \oldbibliography{#1}%
  \setlength{\itemsep}{0pt}%
  \setlength{\parsep}{0pt}%
  \setlength{\parskip}{0pt}%
  \setlength{\bibsep}{0ex}
  \raggedright
}
\newcommand{\photoz}{photo-$z$}
\newcommand{\photozs}{photo-$z$s}

\defcitealias{simbig_letter}{H22a}
\defcitealias{simbig_mock}{H23}

\let\oldAA\AA
\renewcommand{\AA}{\text{\normalfont\oldAA}}

\newcommand{\bitem}{\begin{itemize}}
\newcommand{\eitem}{\end{itemize}}
\newcommand{\beq}{\begin{equation}}
\newcommand{\eeq}{\end{equation}}

\definecolor{orange}{rgb}{1,0.5,0}

\begin{document} \sloppy\sloppypar\frenchspacing 

\title{Inhomogeneous Dust Biases Photometric Redshifts and Stellar Masses for LSST}
\newcounter{affilcounter}
\author[0000-0003-1197-0902]{ChangHoon Hahn}
\altaffiliation{changhoon.hahn@princeton.edu}
\affil{Department of Astrophysical Sciences, Princeton University, Princeton NJ 08544, USA} 

\author[0000-0002-8873-5065]{Peter Melchior}
\affil{Department of Astrophysical Sciences, Princeton University, Princeton NJ 08544, USA} 
\affil{Center for Statistics and Machine Learning, Princeton University, Princeton, NJ 08544, USA}

\begin{abstract}
Spectral energy distribution (SED) modeling is one of the main methods 
to estimate galaxy properties, such as photometric redshifts, $z$, and 
stellar masses, $M_*$, for extragalactic imaging surveys.
SEDs are currently modeled as light from a composite stellar population 
attenuated by a uniform foreground dust screen,
despite evidence from simulations and observations that find large
spatial variations in dust attenuation due to the detailed geometry of stars 
and gas within galaxies. 
In this work, we examine the impact of this simplistic dust assumption on 
inferred $z$ and $M_*$ for Rubin LSST.
We first construct synthetic LSST-like observations ($ugrizy$ magnitudes) from 
the NIHAO-SKIRT catalog, which provides SEDs from high resolution hydrodynamic
simulations using 3D Monte Carlo radiative transfer.
We then infer $z$ and $M_*$ from the synthetic observations using the PROVABGS 
Bayesian SED modeling framework. 
Overall, the uniform dust screen assumption biases both $z$ and $M_*$ in 
galaxies, consistently and significantly for galaxies with dust attenuation $A_V \gtrsim 0.5$, and likely below.
The biases depend on the orientation in which the galaxies are observed. 
At $z=0.4$, $z$ is overestimated by $\sim$0.02 for face-on galaxies and
$M_*$ is underestimated by $\sim$0.4 dex for edge-on galaxies.
The bias in \photoz~is equivalent to the desired redshift precision level 
of LSST ``gold sample'' and will be larger at higher redshifts. 
Our results underscore the need for SED models with additional flexibility 
in the dust parameterization to mitigate significant systematic biases in 
cosmological analyses with LSST. 
\end{abstract} 

\section{Introduction} \label{sec:intro} 
The next generation of photometric galaxy surveys, e.g., the Vera C. Rubin
Observatory’s Legacy Survey of Space and 
Time~\citep[LSST;][]{lsstsciencecollaboration2009}, 
the ESA {\em Euclid} satellite mission~\citep{laureijs2011}, and 
the Nancy Grace Roman Space Telescope~\citep[Roman;][]{spergel2015}, 
will observe billions of galaxies to probe the nature of dark matter and 
dark energy. 
Given their statistical power, cosmological analyses with these surveys will 
no longer be dominated by statistical uncertainties but rather by systematic 
biases. 
Systematic biases in photometric redshifts (\photozs), in particular,
will be a major limitation for these surveys as most of the primary cosmological 
analyses (weak gravitational lensing, galaxy clusters, and supernova host galaxies) 
rely on accurate \photozs~\citep{ivezic2019}. 

Even for current galaxy surveys, there is already evidence that systematic biases
in \photozs~impact cosmological analyses. 
Recent weak lensing analyses of the Hyper Suprime-Cam (HSC) survey 
Year-3 data~\citep{dalal2023, li2023a, more2023, miyatake2023} account for 
residual systematic \photoz~errors by introducing a parameter that allows the 
entire \photoz~distribution, $n(z)$, to shift. 
Indeed, \cite{miyatake2023} find that redshifts of HSC galaxies are
systematically $\sim$0.05 higher than the \photoz~estimates. 
Moreover, when introducing this parameter, their $S_8$ constraints, a key 
cosmological parameter combination that measures the cosmic growth of structure, 
is shifted from $0.791^{+0.018}_{-0.020}$ to $0.763^{+0.040}_{-0.036}$ 
--- a $\sim$1$\sigma$ shift.

While accounting for \photoz~errors by shifting the entire $n(z)$ distribution 
may be sufficient for current analyses, it ignores the fact that \photoz~errors 
likely depend on galaxy properties. 
Hence, there is no guarantee that it will be sufficient for unbiased cosmological
analyses with upcoming surveys, specifically when galaxies are grouped into various samples.
Furthermore, even if it is sufficient, it roughly {\em halves} the constraining 
power of the  survey. 

Current methods for estimating \photoz~roughly fall into two broad categories. 
The first are data-driven methods that bootstrap spectroscopic galaxy samples 
to learn the mapping between \photozs~and the more accurate and precise 
spectroscopic redshifts. 
The latest methods use self-organizing maps~\citep[e.g.,][]{wright2020}. 
In principle, these methods produce unbiased and robust \photoz~estimates for
galaxies within the selection of the spectroscopic sample.
However, cosmological analyses with these methods are ultimately limited by  
the spectroscopic sample selection, which is almost always much shallower 
than photometric surveys.
A large fraction of the data and constraining power from the
survey hinges on a problematic calibration.

The other class of methods involve modeling of the spectral energy distribution (SED)~\citep[e.g.,][]{brammer2008, tanaka2015}.
Going beyond fitting of spectral templates, current SED models assume that galaxy SEDs arise from composite stellar populations, which are attenuated by 
dust~\citep[see][for a review]{conroy2013}. 
High-resolution galaxy formation simulations with 3D Monte Carlo dust radiative 
transfer (RT) reveal that variations in the ``star-dust geometry'', the spatial 
distribution between stars and the dusty interstellar medium (ISM)
produce significant spatial variation in dust attenuation~\citep{narayanan2018a}. 
Observations have corroborated this for decades with the diversity of
measured attenuation curves~\citep[e.g.,][]{cardelli1989, pei1992, calzetti2000, conroy2010a, chevallard2013, reddy2015}.
Even by eye, one easily finds dust lanes and other complex star-dust geometry
within galaxies in high-resolution images. 
Yet most state-of-the-art models assume a simple dust geometry: a uniform 
foreground dust 
screen~\citep[e.g.,][]{dacunha2011, boquien2019, prospector,  hahn2023}.

In this work, we examine whether this simplistic dust assumption biases 
inferred \photozs~in upcoming cosmological galaxy surveys. 
We also consider stellar masses, $M_*$, another key galaxy property for 
cosmological analyses used to characterize how galaxy samples trace the 
underlying matter distribution.
We focus on LSST and construct synthetic LSST-like observations from the NIHAO-
SKIRT~\citep{faucher2023}, a catalog of SEDs constructed from 65 galaxies in 
a hydrodynamical simulation using stellar population synthesis (SPS) and 3D 
Monte Carlo RT. 
Then, we apply a state-of-the-art SED modeling~\citep[PROVABGS;][]{hahn2023} 
to infer $z$ and $M_*$ and quantify their biases. 

We begin in Section~\ref{sec:nihao} with a brief explanation of our synthetic 
LSST photometry. 
We then describe the SED modeling framework used to infer \photoz~and 
$M_*$ in  Section~\ref{sec:provabgs}. 
Afterwards, we present the results and discuss their implications in 
Section~\ref{sec:results}. 
\begin{figure*}
    \includegraphics[width=\textwidth]{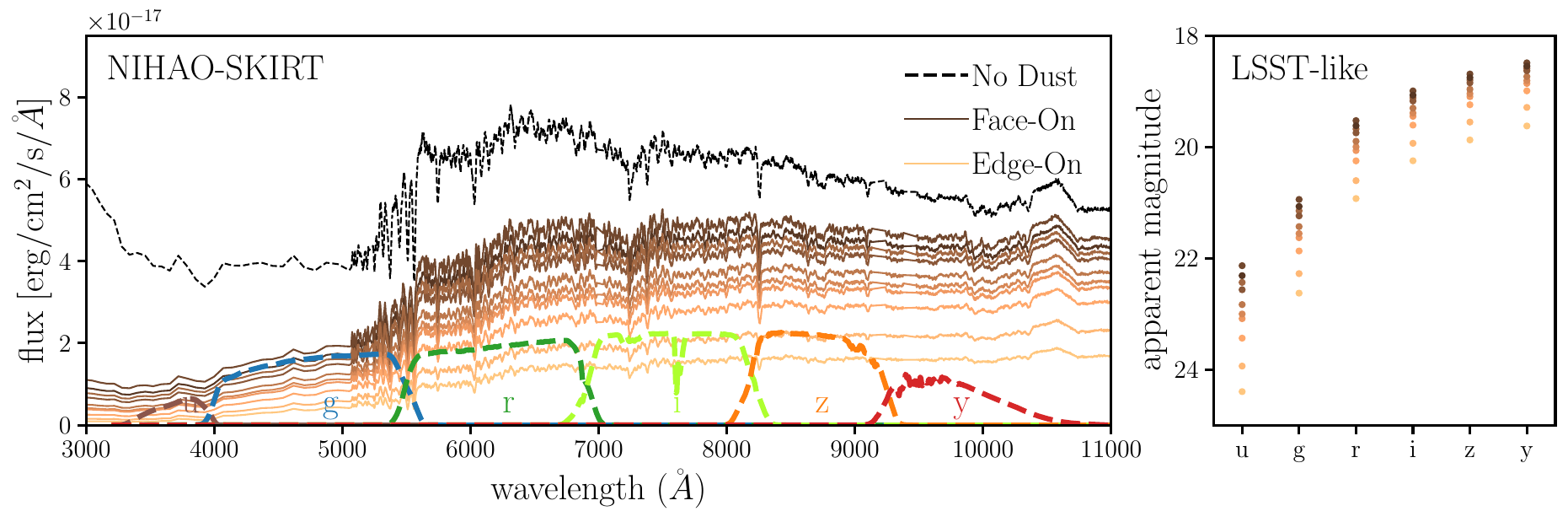}
    \caption{\label{fig:lsst}
    {\em Left panel}: SEDs from the NIHAO-SKIRT catalog for one of the 
    simulated galaxies. 
    We include the SEDs of the galaxy observed from 10 different orientations 
    ranging from  face-on to edge-on (color gradient). 
    We also include an SED excluding dust (black dashed). 
    The SEDs are modeled at $z=0.4$ and their emission lines are masked. 
    For reference, we plot the LSST broadband throughput curves (dotted). 
    {\em Right panel}: Synthetic LSST magnitudes constructed from the 
    NIHAO-SKIRT SEDs observed from the 10 different orientations. 
    The magnitudes include LSST-like observational uncertainties. 
    }
\end{figure*}

\section{NIHAO-SKIRT Catalog} \label{sec:nihao}
We construct our synthetic LSST observations using the NIHAO-SKIRT 
catalog~\citep{faucher2023}. 
NIHAO-SKIRT provides the SEDs of 65 galaxies from the Numerical Investigation 
of Hundred Astrophysical Objects (NIHAO) hydrodynamical simulation
suite~\citep{wang2015}. 
The SEDs are constructed using stellar population synthesis (SPS) with
FSPS~\citep{conroy2009, conroy2010, foreman-mackey2023} and 3D Monte Carlo 
radiative transfer using SKIRT~\citep{camps2020}. 
The SPS assumes a~\cite{chabrier2003} stellar initial mass function and
uses the MIST isochrones~\citep{paxton2011, paxton2013, paxton2015, choi2016, dotter2016} 
and the MILES spectral library~\citep{sanchez-blazquez2006}.
The dust attenuation and emission from diffuse interstellar medium (ISM)
is then calculated with SKIRT, which uses the THEMIS dust modeling framework~\citep{jones2017}.
For young stellar populations, \cite{faucher2023} uses a subgrid recipe because 
photodisassociation regions are unresolved in the NIHAO galaxies.
NIHAO-SKIRT is based on multiphase ISM and, thus, has more realistic dust density
structure and diverse attenuation curves in comparison to previous works by,
e.g., \cite{hayward2015, lower2020, trcka2020}.
In total, the NIHAO-SKIRT catalog provide 650 SEDs of 65 NIAHO galaxies each
observed from 10 different orientations that range from roughly face-on to
roughly edge-on. 
It also includes matching SEDs that exclude any effect of dust. 

We construct our synthetic LSST observations from the NIHAO-SKIRT SEDs, which 
are modeled at a distance of 100 Mpc. 
First, we ``dry-redshift'' the SEDs to $z=0.4$ --- i.e., we redshift the 
wavelength of the SED and scale fluxes by the new luminosity distance. 
We then mask emission line of the SEDs and convolve them with the LSST 
photometric broadband filters  in the $ugrizy$-bands. 
We mask the emission lines to mitigate any model misspecifications: 
SPS does not include the contributions from nebular gas and the PROVABGS SED 
model does not model emission lines. 
Lastly, we incorporate realistic LSST-like observational uncertainties to the 
modeled photometric magnitudes.
We apply Gaussian noise with a magnitude-dependent amplitude. 
More specifically, we use the magnitude error versus apparent magnitude relation
in \cite{graham2018}, which is constructed using the \cite{connolly2014} LSST 
simulation package.

In Figure~\ref{fig:lsst}, we present the NIHAO-SKIRT SEDs (left panel) and 
the synthetic LSST magnitudes constructed from them (right panel) for a single 
simulated NIHAO galaxy. 
In the left panel, we present the SEDs of the galaxy observed from 10 different 
orientations.
The color gradient represent the orientation ranging from edge-on to face-on: 
light to dark. 
We also include an SED excluding any effects of dust (black dashed), observed
edge-on. 
The SEDs are modeled at $z=0.4$ and their emission lines are masked. 
For reference, we plot the LSST broadband throughput curves (dotted). 
In the right panel, we present the synthetic LSST magnitudes for the 10 
different orientations. 
The error bar represent the LSST-like observational uncertainties that we include; however, 
they are negligible in the figure. 

\section{SED modeling with PROVABGS} \label{sec:provabgs}
We infer \photoz~and $M_*$ from the synthetic LSST magnitudes using 
the PROVABGS Bayesian SED modeling framework~\citep{hahn2023, kwon2023}. 
PROVABGS models galaxy SEDs using SPS based on FSPS with MIST isochrones, MILES 
stellar libraries, and the \cite{chabrier2003} IMF. 
These are the same modeling choices as the SPS in NIHAO-SKIRT. 
Furthermore, PROVABGS uses a non-parametric star-formation history (SFH) 
with a starburst and a non-parametric evolving metallicity history (ZH). 
These prescriptions are derived from SFHs and ZHs of simulated galaxies in the
Illustris hydrodynamic simulation~\citep{vogelsberger2014, genel2014, nelson2015} 
and provide compact and flexible representations of a wide range of SFHs and ZHs. 

For dust attenuation, PROVABGS uses the two-component \cite{charlot2000} model with 
birth cloud and ISM components.
Both components are modeled as a uniform foreground dust screen that attenuates
all of the star light from a galaxy. 
We use PROVABGS primarily for its speed and convenience: e.g., PROVABGS enables
accelerated inference through neural emulators for the SED model~\citep{kwon2023}.

For the Bayesian inference, we estimate the posteriors using the Preconditioned 
Monte Carlo algorithm $\mathtt{pocoMC}$\footnote{\url{https://pocomc.readthedocs.io/en/latest/}}~\citep{karamanis2022accelerating, karamanis2022pocomc}.
In total the PROVABGS SED model that we use has 13 parameters: \photoz, $M_*$, 
6 parameters specifying the SFH 
($\beta_1, \beta_2, \beta_3, \beta_4, f_{\rm burst}, t_{\rm burst}$), 
2 parameters specifying ZH ($\gamma_1, \gamma_2$), and 3 parameters
specifying dust attenuation ($\tau_{\rm BC}, \tau_{\rm ISM}, n_{\rm dust}$). 
PROVABGS infers full posteriors over all 13 parameters, but for this work 
we focus solely on the posteriors of $z$ and $M_*$.
To emphasize the critical point of this paper: Our SED model does include 
dust attenuation from birth clouds and ISM but, like other current SED models
(e.g., Prospector,~\citealt{prospector}; CIGALE,~\citealt{noll2009a, boquien2019}), 
it treats dust as a homogeneous screen, whereas galaxies show highly 
inhomogeneous dust.
We want to know how this model misspecification affects SED parameter 
constraints on $z$ and $M_*$.

\begin{figure*}
\begin{center}
    \includegraphics[width=0.5\textwidth]{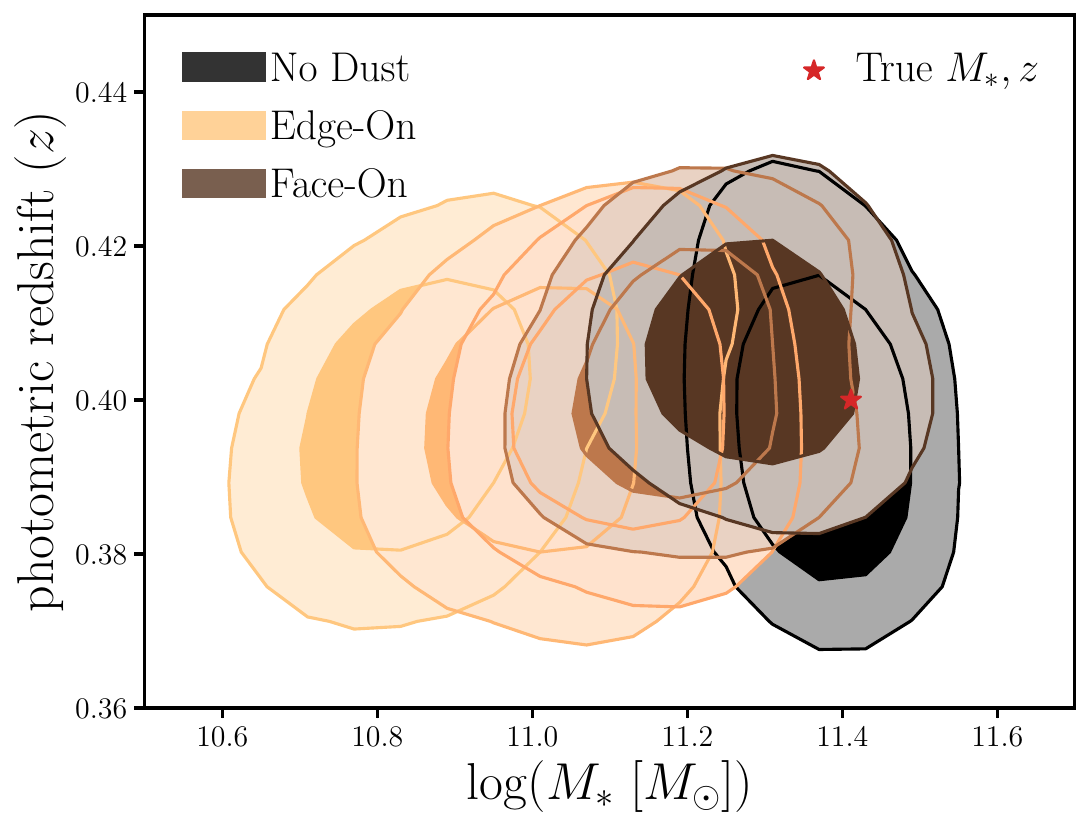}
    \caption{\label{fig:posterior}
    Posteriors, $p(z, M_*\,|\,{\bf X}_{i,j})$, for a single simulated 
    NIHAO-SKIRT galaxy ($i=64$) observed from different orientations, $j$.
    We present five of the ten orientations ranging from edge-on to face-on 
    (light to dark). 
    We also include the posterior inferred from ${\bf X}_{i,j}^{\rm no\,dust}$, 
    LSST photometry constructed from SEDs with no dust (black), as well as 
    the true $z$ and $M_*$ (red star). 
    {\em Dust introduces orientation dependent biases in the inferred $z$ and 
    $M_*$.}
    }
\end{center}
\end{figure*}

\section{Results \& Discussion} \label{sec:results}
We use PROVABGS to infer the posteriors of $z$ and $M_*$ from the LSST-like 
magnitudes with and without dust, ${\bf X}^{\rm dust}_{i,j}$ and 
${\bf X}^{\rm no\,dust}_{i,j}$, for $i = 1, ... 65$ NIHAO galaxies observed 
at $j = 1, ... 10$ orientations.
In Figure~\ref{fig:posterior}, we present the posteriors for a single 
massive galaxy ($i=64$, $M_{*,i} = 10^{11.41} M_\odot$) at different 
orientations ranging from edge-on to face-on (light to dark): 
$p(z, M_*\,|\,{\bf X}^{\rm dust}_{64, j})$. 
We select five out of ten orientations for clarity. 
For reference, we include $p(z, M_*\,|\,{\bf X}^{\rm no\,dust}_{64,j})$, 
the posterior inferred using the magnitudes without  any dust (black). 
We also mark the true $z$ and $M_*$ of the galaxy (red). 

For all galaxies, the posteriors have consistent precision on the $z$ and $M_*$ 
constraints: $\sigma_z \sim 0.01$ and $\sigma_{\log M_*} \sim 0.07$.
Their accuracy, however, significantly differ. 
First, for photometry without dust, we are able to infer a posterior that 
is fully consistent with the true $z$ and $M_*$. 
Meanwhile, the posteriors for photometry with dust shift significantly. 
For \photoz, the inferred $z$ has a noticeable dependence on galaxy 
orientation: we infer higher $z$ for more face-on orientations. 
For the most face-on orientation, the inferred $z$ is slightly higher 
than the true $z$ --- though statistically consistent. 
For $M_*$, the inferred values are increasingly underestimated for more 
edge-on orientations.
The trend is significant and monotonic.  
For the most edge-on orientation, the inferred $M_*$ is more than 0.5 dex 
lower than the true value --- a discrepancy of >4$\sigma$. 

\begin{figure*}
\begin{center}
    \includegraphics[width=0.9\textwidth]{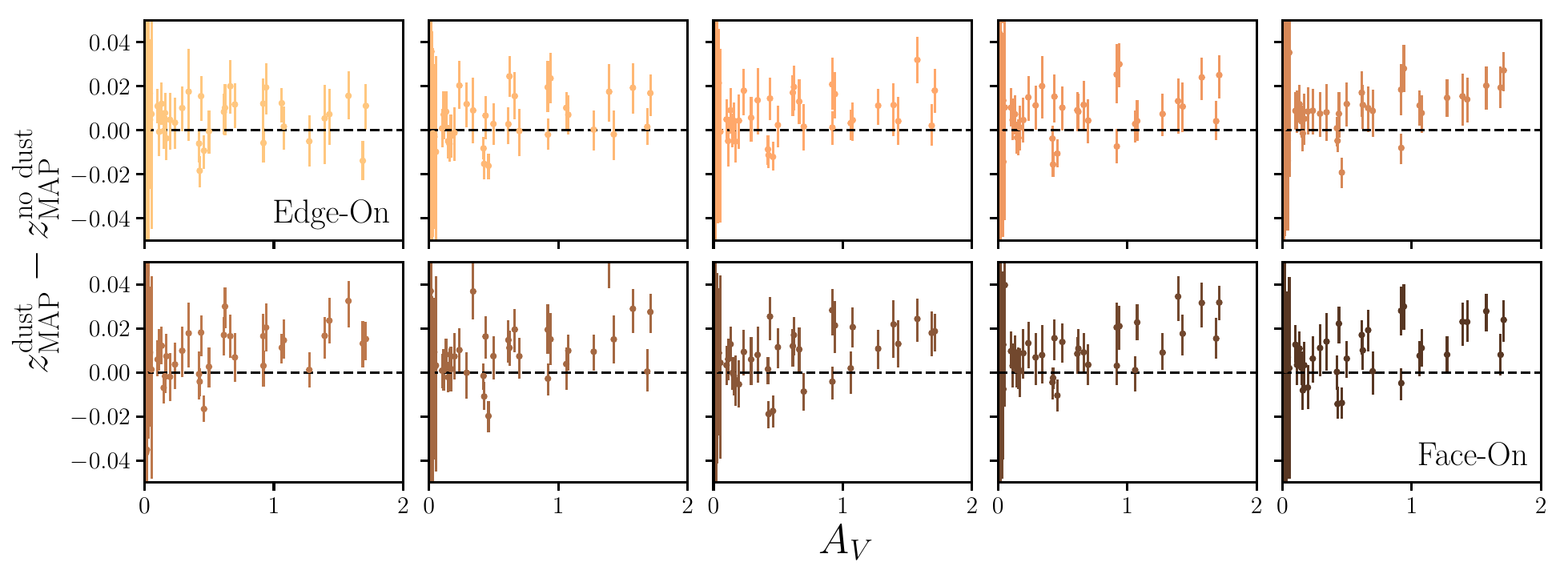}
    \includegraphics[width=0.9\textwidth]{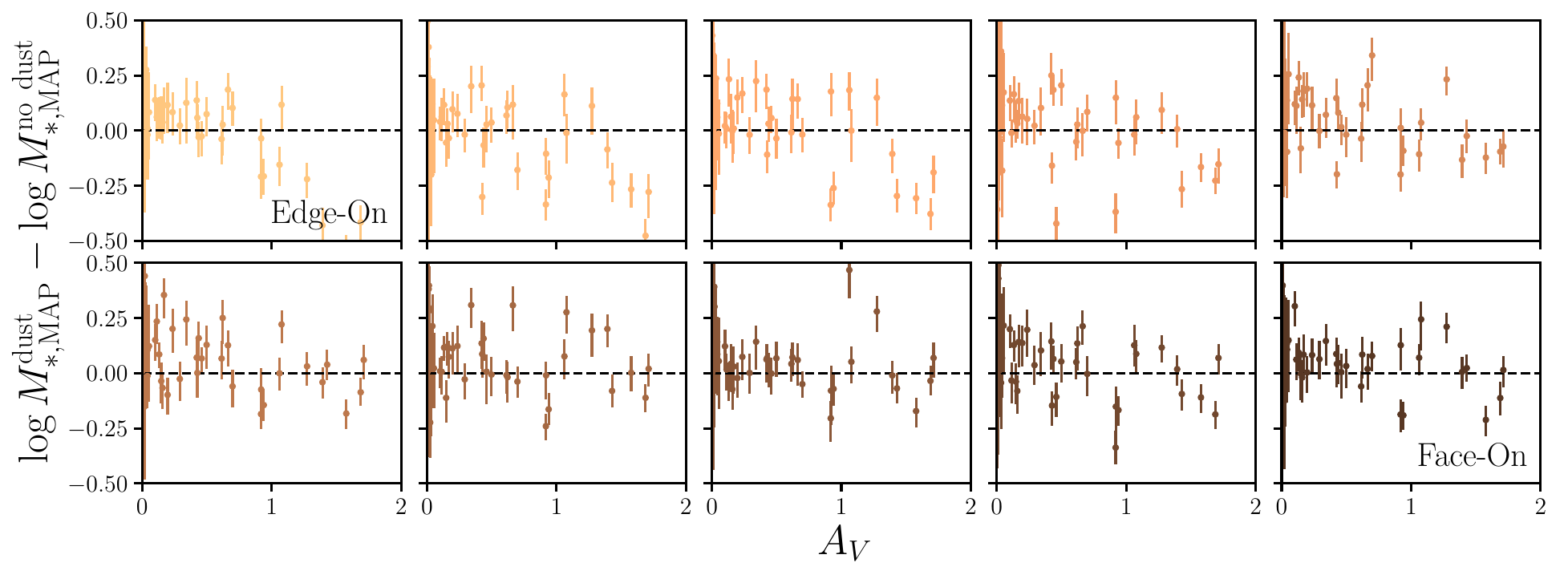}
    \caption{\label{fig:residual}
    The effect of dust on inferred \photoz~(top) and $M_*$ (bottom). 
    We present the difference between the MAP values of 
    $p(z, M_*\,|\,{\bf X}^{\rm dust}_{i,j})$ and 
    $p(z, M_*\,|\,{\bf X}^{\rm no\,dust}_{i,j})$ as a function of $A_V$. 
    Each panel and color presents a different orientation from edge-on
    (top left; light) to face-on (bottom right; dark). 
    With $A_V \lesssim 0.5$ we infer overall unbiased $z$ and $M_*$. 
    With $A_V > 0.5$, there is a bias in \photozs~that depends on orientation:
    \photozs~are overestimated for more face-on orientations.
    For face-on, \photozs~are overestimated by $\sim$0.02.  
    There is also a bias in $M_*$ that depends on orientation: $M_*$ is
    underestimated for more edge-on orientations.
    For edge-on, $M_*$ are underestimated by $\sim$0.4 dex.  
    The current uniform dust screen assumption in SED modeling is insufficient
    at marginalizing the effect of dust and will impact LSST cosmological analyses.}
\end{center}
\end{figure*}

We examine the role of dust and galaxy orientation further, in 
Figure~\ref{fig:residual}. 
We compare the posteriors from photometry with dust versus without dust for
\photoz~(top) and $M_*$ (bottom) as a function of dust attenuation in the 
$V$-band, $A_V$.
In particular, we present the differences between the maximum-a-posterior 
(MAP) values of $p(z, M_*\,|\,{\bf X}^{\rm dust}_{i,j})$ and of
$p(z, M_*\,|\,{\bf X}^{\rm no\,dust}_{i,j})$ for all 65 NIHAO galaxies. 
The errorbars represent the 16 and 84$^{th}$ percentiles of 
$p(z, M_*\,|\,{\bf X}^{\rm dust}_{i,j})$, for reference. 
Each panel and color presents the differences for a single galaxy orientation 
ranging from edge-on (top left; light) to face-on (bottom right; dark). 

We compare $p(z, M_*\,|\,{\bf X}^{\rm dust}_{i,j})$ against
$p(z, M_*\,|\,{\bf X}^{\rm no\,dust}_{i,j})$, instead of against the true 
$z$ and $M_*$, to isolate the effects of dust from the broader model 
misspecification issues in SED modeling. 
Although we find overall good consistency between 
$p(z, M_*\,|\,{\bf X}^{\rm no\,dust}_{i,j})$ and the true values,
\cite{faucher2024} recently find that model misspecification can significantly 
biases, e.g., inferred SFRs by more than a factor of 3. 
In this work, we solely focuses on the impact of dust in biasing 
\photoz~and $M_*$; we refer readers to \cite{faucher2024} for a broader 
examination of model misspecification in SED modeling. 

For the NIHAO galaxies with higher dust attenuation ($A_V > 0.5$), Figure~\ref{fig:residual} reveals a significant 
an orientation dependent bias, consistent with Figure~\ref{fig:posterior}. 
For \photoz, more face-on orientations overestimate $z$.
\photozs~are overall unbiased for the edge-on orientation but overestimated
by $\sim$0.02 for the face-on orientation. 
This corresponds to roughly a 2$\sigma_z$ bias.
It is also equivalent to the desired precision level of the 
high signal-to-noise LSST ``gold sample'', which will be used for the majority
of cosmological analyses~\citep{lsstsciencecollaboration2009}. 
For lower dust attenuation ($A_V \lesssim 0.5$), the results 
are inconclusive given the photometric errors of individual 
galaxies, but are likely still biased when aggregating larger 
samples.

There is also an orientation dependent bias for $M_*$, where $M_*$ is 
significantly underestimated for more edge-on orientations. 
The inferred $M_*$ values are overall unbiased for the face-on orientation but
underestimated by $\sim$0.4 dex for the edge-on orientation. 
This corresponds to a $>$4$\sigma_{M_*}$ bias. 
We note that this bias in $M_*$ is consistent with the $M_*$ bias found in 
\cite{faucher2024} using different SED modeling methods. 

The biases in $z$ and $M_*$ come from the fact that dust significantly 
impacts the photometry of galaxies, but the parameters of a SED model that 
best fit the attenuated photometry are inconsistent with the ones that best 
fit the unattenuated photometry. 
In other words, the dust attenuation model in our SED model does not 
sufficiently marginalize over the effect of dust.
Hence, the assumption of a uniform dust screen is insufficient for the 
precision level of LSST photometry.
Interestingly, we find that the uniform dust screen assumption breaks down even 
for face-on galaxies. 
Edge-on galaxies will generally have significantly more attenuation than
face-on galaxies, as stellar light will need to travel through more of
the dust-filled ISM. 
Nevertheless, \photozs~are biased for face-on galaxies. 

The bias in \photozs~has significant implications for LSST cosmological 
analyses. 
As mentioned above, biases in \photoz~measurements of source galaxies in 
weak lensing analyzes significantly bias $S_8$ 
constraints~\citep[e.g.,][]{more2023}. 
We find significant \photoz~bias for galaxies with $A_V > 0.5$.
In our NIHAO sample this consists of massive galaxies with $M_* > 10^{10}M_\odot$. 
For LSST weak lensing galaxies, the  sample will include galaxies brighter than
$i < 25.3$ spanning 
$0.35 \lesssim z \lesssim 2$~\citep{lsstsciencecollaboration2009}. 
Given the selection and wide redshift range, we expect that the impact of dust will be more 
significant than our findings at $z=0.4$ and show a significant redshift 
dependence. 

First, with the magnitude cut, the fraction of massive galaxies will increase
with redshift. 
Since more massive galaxies generally have more 
dust~\citep[e.g.,][]{ciesla2014, remy-ruyer2014, santini2014}, 
the impact of dust in biasing \photozs~will also increase with redshift. 
The overall \photoz~bias will also be made worse due to differences in the 
properties of galaxies at higher redshift. 
Higher redshift galaxies have higher dust fractions. 
For instance, \cite{santini2014} find, for a galaxy sample with FIR observations, 
that at fixed $M_*$ galaxies at $z\sim 2$ have $\sim10$ times higher dust 
fraction than galaxies at $z\sim0.4$. 
Thus, we can expect dust to bias \photozs~at even lower masses at higher 
redshifts. 

Beyond the increased bias in \photozs, the redshift dependence of the increase 
implies that bias will not simply shift the overall $n(z)$. 
Since higher-redshift galaxies will be more biased, the bias will change the 
overall shape of $n(z)$. 
Therefore, introducing a parameter that shifts $n(z)$, as in the HSC analyses,
will not fully account for this effect for LSST  weak-lensing analyses. 

The bias in $M_*$ also has significant implications for LSST. 
$M_*$ is a key physical property used to study galaxy evolution and
measure, e.g., galaxy mass assembly. 
Furthermore, it plays a critical role in understanding the galaxy-halo 
connection through galaxy bias or abundance matching~(see \citealt{wechsler2018} 
for a review). 
For cosmological analyses, the galaxy-halo connection is a key component in 
construct large-volume simulations for estimating the covariance matrices 
of observables and validating the analyses. 
Since dust leads to systematic underestimates of $M_*$, it can lead us to 
mischaracterize the galaxy-halo connection. 

Fortunately, it is possible to mitigate the impact of dust on the inferred \photozs~and $M_*$. 
The main issue stems from model misspecification caused by the uniform 
dust screen assumption in SED models.
We therefore recommend the adoption of non-uniform dust screens that 
include multiple dust components~\citep[e.g.,][]{lower2022}, which will 
provide a more flexible description of a wider range of dust attenuation 
curves and, thus, enable us to better marginalize over the effect of dust. 
One consequence of this approach, or any increase in the flexibility of the 
model, is that the constraints on \photozs~and $M_*$ will get weaker.
In future work, we will investigate ways to determine spatial dust 
distributions to provide unbiased SED-derived parameters, regain some 
or all of the constraining power of photometric survey data, and 
study the relation between stellar and dust morphologies of galaxies.

\section*{Acknowledgements}
It's a pleasure to thank Nicholas Faucher and Michael R. Blanton for 
useful discussions and for publicly releasing the NIHAO-SKIRT catalog.
This work was supported by the AI Accelerator program of the Schmidt Futures Foundation. 
This work was substantially performed using the Princeton Research Computing resources
at Princeton University, which is a consortium of groups led by the Princeton Institute 
for Computational Science and Engineering (PICSciE) and Office of Information Technology’s 
Research Computing.

\bibliography{dusty} 
\end{document}